\documentclass[twocolumn,tighten,times,tracking]{aastex}

\usepackage{amsmath}
\usepackage{color}
\usepackage{longtable}
\usepackage{ulem,array}
\usepackage{grffile}
\usepackage{comment}
\usepackage{hyperref}
\hypersetup{
pageanchor=false,
colorlinks=true,
linkcolor=blue,
citecolor=blue
}

\usepackage{array}
\usepackage{tabularx} 
\newcolumntype{Y}{>{\centering\arraybackslash}X} 

\newcolumntype{M}[1]{>{\centering\arraybackslash}m{#1}}
\newcolumntype{N}{@{}m{0pt}@{}}

\newcommand{\mathbfit}[1]{\textbf{\textit{#1}}}
\renewcommand{\vec}[1]{\mathbfit{#1}}

\newcommand{\alf}{Alfv$\acute{\text{e}}$n}

\newcommand{\Ma}{\mathcal{M}_A}
\newcommand{\Ms}{\mathcal{M}_{\rm s}}

\usepackage{multirow}
\usepackage{amssymb}

\makeatletter
\newcommand{\xRightarrow}[2][]{\ext@arrow 0359\Rightarrowfill@{#1}{#2}}
\makeatother

\usepackage[mathscr]{euscript}

\newcommand{\sig}{\sigma_i}
\newcommand{\sigd}{\sigma_i^{\prime}}
 
\newcommand{\add}{\textcolor{black}}

\DeclareSymbolFont{matha}{OML}{txmi}{m}{it}
\DeclareMathSymbol{\varv}{\mathord}{matha}{29}

\SetSymbolFont{symbols}{bold}{OMS}{cmsy}{b}{n}
\DeclareSymbolFont{bmisymbols}{OML}{cmm}{b}{it}
\DeclareMathSymbol{\bvarv}{0}{bmisymbols}{"1D}

\begin{document}

\author[0000-0001-9625-5929]{Mohamad Shalaby}
\altaffiliation{Horizon AstroPhysics Initiative (HAPI) Fellow}
\affiliation{Perimeter Institute for Theoretical Physics, 31 Caroline Street North, Waterloo, ON, N2L 2Y5, Canada;}
\affiliation{Waterloo Centre for Astrophysics, University of Waterloo, Waterloo, ON N2L 3G1, Canada;
}
\affiliation{Department of Physics and Astronomy, University of Waterloo, 200 University Avenue West, Waterloo, ON, N2L 3G1, Canada
}

\author[0000-0003-2030-0046]{Antoine Bret}
\affiliation{
ETSI Industriales, Universidad de Castilla-La Mancha, 13071 Ciudad
Real, Spain;}
\affiliation{Instituto de Investigaciones Energéticas y Aplicaciones Industriales,
Campus Universitario de Ciudad Real, 13071 Ciudad Real, Spain}

\author[0000-0002-5456-4771]{Federico Fraschetti}
\affiliation{Center for Astrophysics | Harvard \& Smithsonian, Cambridge, MA, 02138, USA;}
\affiliation{Department of Planetary Sciences—Lunar and Planetary Laboratory, University of Arizona, Tucson, AZ, 85721, USA}

\title{
Parallel Collisionless Shocks in strongly Magnetized Electron-Ion Plasma.
I. Temperature anisotropies
}
\shorttitle{ }
\shortauthors{Shalaby et al.}

\begin{abstract}
    Collisionless electron-ion shocks are fundamental to astrophysical plasmas, yet their behavior in strong magnetic fields remains poorly understood. Using Particle-in-Cell (PIC) simulations with the SHARP-1D3V code, we investigate the role of the ion magnetization parameter $\sigma_i$ in parallel shock transitions. Strongly magnetized converging flows ($\sigma_i > 1$) exhibit lower density compression ratios ($R \sim 2$), smaller entropy jumps, and suppressed particle acceleration, while maintaining pressure anisotropy stability due to conserved perpendicular temperatures across the transition region,
    alongside increased parallel temperatures. In contrast, weakly magnetized shocks drive downstream mirror and firehose instabilities due to ion temperature anisotropy, which are suppressed in strongly magnetized cases. Additionally, weakly magnetized shocks exhibit the onset of a supra-thermal population induced by shock-drift acceleration, with most of the upstream kinetic energy thermalized for both electrons and ions in the downstream region. Our results demonstrate that perpendicular temperatures for both species are conserved in \add{weakly and }strongly magnetized cases and highlight deviations from standard ideal magnetohydrodynamic (MHD) behavior \add{in strongly magnetized cases}. These findings provide critical insights into the role of magnetic fields in parallel collisionless astrophysical shocks.
\end{abstract}

\section{Introduction}
\label{sec:intro}
Shock waves in fluids are fundamental phenomena that have been studied for over 100 years \citep{tidman1971,Whitham:74,Zeldovich,LandauFluid,Salas2007}. In astrophysical context, shock waves are often collisionless, i.e. they propagate in a medium where the mean free path is much greater than the shock front \add{thickness} \citep{Balogh2013,Burgess2015}. Since in a fluid, binary collisions mediate the transition between the upstream and the downstream, the absence of such collisions long cast some doubts on the very existence of collisionless  shock waves \citep{Sagdeev_Kennel_1991}, the study of which dates back to the 1960s with the pioneering work of Sagdeev \citep{Sagdeev66}. Their reality has now been established for decades thanks, for example, to \textit{in situ} observations of the bow shock of the Earth's magnetosphere in the solar wind \citep{PRLBow1,PRLBow2}. 

It is now well-established that shocks are ubiquitous in astrophysics, ranging from interplanetary shocks in the solar system to supernova remnants \citep{Hollenbach1980}. In-situ measurements at Earth's bow shock have revealed temperature anisotropies in the downstream plasma, which are known to trigger firehose instabilities at quasi-parallel shocks and mirror instabilities at quasi-perpendicular shocks \citep{Bale.etal:09,MarucaPRL2011,SchlickeiserPRL2011}. For instance, Wind measurements \citep{Vogl.etal:01} during an oblique crossing of Earth's bow shock show $\sim \% 1$ deviations in density compression, which were explained therein by pressure anisotropies. Similarly, Voyager 1 data from the quasi-perpendicular solar wind termination shock crossing provided evidence of mirror instabilities in downstream region \citep{Liu.etal:07,Genot:08}. 

A statistical analysis of 91 interplanetary Coronal Mass Ejection (ICME) sheaths using Wind data has demonstrated that the Alfvén Mach number is a key parameter governing mirror-mode instabilities at interplanetary shocks \citep{Ala-Lahti.etal:18}. Such magnetic fluctuations are likely driven by temperature anisotropies generated by instabilities near the shock. Therefore, a kinetic-scale analysis of the relationship between the Mach number and temperature anisotropy can provide valuable insights for interpreting such observations.

In contrast, temperature anisotropies at quasi-parallel shocks are more challenging to study numerically and model due to the significantly higher level of upstream fluctuations compared to quasi-perpendicular shocks. As a result, fewer studies have been reported on this topic \citep{Fuselier.etal:94,Rojas.etal:23}.

Collisionless shocks are often studied using the magnetohydrodynamic (MHD) formalism, although this formalism is ultimately based on fluid mechanics, which assumes a small mean free path.
Global magnetohydrodynamic (MHD) simulations suggest that mirror-mode-dominated turbulence may be prevalent in the inner heliosheath behind the termination shock \citep{Fichtner.etal:20}, potentially influencing the transport of galactic cosmic rays as they enter the heliosphere.
It is therefore important to consider the extent to which such an approach is justified when one of its basic hypothesis is no longer valid.

In general, two potential sources of divergence between the behavior of a collisionless shock and that of an MHD shock, have been identified in recent years (see \cite{Bret2020} and references therein). First, in the absence of collisions, a magnetized plasma can be stable even if anisotropic \citep{Gary1993}.
Yet, MHD jump equations consider, at least in their simplest version, an isotropic medium. Secondly, it is known that collisionless shocks can accelerate particles \citep{Axford1977,1987Blandford}, which is the primary reason for their significant interest in astrophysics, as they may partly explain the origin of cosmic rays.

However, these accelerated particles escape the so to speak ``Rankine Hugoniot budget’’ on which the MHD jump conditions are based \citep{Berezhko1999ApJ,David.etal:22}. This budget assumes that all of the upstream fluid passes into the downstream, together with the matter, momentum, and energy it carries. If some of it, the cosmic rays, moves back and forth around the shock front or returns from the downstream to the upstream, this assumption of integral transport from the downstream to the upstream is no longer valid.

The present article deals with the first cause of divergence, namely the possibility that an anisotropy developing at the front crossing remains stable in the downstream. In the absence of an ambient magnetic field, such a configuration would return to isotropy on a time scale given by the Weibel instability \citep{Weibel1959,SilvaPRE2021}. However, in the presence of an ambient magnetic field, an anisotropic configuration can be stable.

In the case of a parallel collisionless shock, i.e. a shock that propagates parallel to the ambient magnetic fields, \cite{Bret+Ramesh2018} developed a model capable of accounting for these phenomena. This model considered the simple case of a shock in a pair plasma, which makes it possible to treat the parallel and perpendicular temperatures of electrons and positrons on an equal footing. Confirmed by Particle-in-Cell (PIC) simulations \citep{Haggerty+2022}, the main conclusion is that a strong parallel magnetic field reduce the density ratio $R$ of a strong sonic shock from 4 to 2. Such a reduction in the density jump would have significant effects on the production of cosmic rays since the power-law index of their momentum spectrum is $3R/(R-1)$ \citep{Blandford1978}.

Further studies varying field obliquity showed that the departure from MHD is indeed maximum for parallel shocks and minimum for perpendicular shocks \citep{BretPoP2019,BretJPP2024}. It is therefore relevant to emphasize parallel shocks. In particular, while the model of \cite{Bret+Ramesh2018} and the PIC simulations confirming it \citep{Haggerty+2022} considered pair plasmas, it is important to assess the case of electrons/ions plasmas. Such is the aim of the present article.

The magnetohydrodynamic (MHD) jump conditions for anisotropic plasmas are derived under the assumption of a bi-Maxwellian velocity distribution \citep[in both the parallel and perpendicular directions to the magnetic field,][]{Hudson1970}. However, this assumption can only be validated through kinetic-scale simulations. Over the past decades, Particle-in-Cell (PIC) simulations have enabled the separation of ion-scale and electron-scale processes \citep[see, e.g.,][]{Pohl.etal:20}. Although the total plasma pressure is dominated by ion pressure, electron-scale instabilities can only be distinguished from ion-scale phenomena with realistic scale separation.

Pristine solar wind has been observed to exhibit increasing temperature anisotropy closer to the Sun. Initial measurements of proton temperature anisotropy by Wind at $1$ au \citep{Bale.etal:09} were later confirmed by Parker Solar Probe (PSP) \add{\citep{Huang.etal:20,Huang.etal:25}}, which revealed even more extreme anisotropy at smaller heliocentric distances (i.e., lower plasma beta), at $< 0.25$ au. To distinguish shock-induced temperature anisotropy from instabilities inherent to the quiescent solar wind, we perform a kinetic-scale analysis of the former for a parallel shock propagating in a laminar medium. However, the transition layer rapidly develops magnetic field fluctuations, leading to local deviations from the parallel geometry, as observed by PSP \citep[e.g.,][]{Jebaraj.etal:24}. A significant consequence of \add{these instability-related, but also pre-existing,} fluctuations is their impact on particle acceleration, as demonstrated by test particle simulations \citep[e.g.,][]{Fraschetti.Giacalone:15}.

The paper is organized as follows. In Section~\ref{sec:theory}, we discuss the theoretical expectations for departures from standard MHD shock behavior. The numerical setup for our PIC electron-ion non-relativistic shock simulations is described in Section~\ref{sec:setup}. Section~\ref{sec:results} explores the effects of strong magnetic fields in electron-ion shock simulations on shock jump conditions, including density, entropy, and temperatures. We demonstrate that strongly magnetized shocks can maintain pressure anisotropy stability in the downstream region, leading to low compression ratios ($\sim 2$) and preventing particle acceleration in these simulations. Finally, we conclude and summarize our findings in Section~\ref{sec:conclusions}. Throughout this work, we use the SI system of units.

\section{Theoretical considerations}
\label{sec:theory}

Let us briefly recall the outline of the model presented in \cite{Bret+Ramesh2018}. The MHD jump equations for non-isotropic temperatures were derived long ago \citep{Hudson1970}. However, in these equations, the degree of anisotropy remains a free parameter that MHD itself cannot determine. Starting from an upstream region assumed to be isotropic, the main idea is to provide a means of determining the downstream anisotropy.

To this end, we reason that when the plasma is compressed across the shock front, the compression is not isotropic but rather takes place in the direction perpendicular to the displacement. As a result, immediately after the shock front, the downstream region experiences an increase in its parallel temperature, while its perpendicular temperature remains unchanged. The resulting degree of anisotropy can be stable if the magnetic field is strong enough. Otherwise, the downstream region migrates towards the threshold of its instability, i.e., the firehose instability.

In a pair plasma, and in the context of collisionless shocks, the magnetic field is characterized by the well-known $\sigma$ parameter \citep{Sironi2011ApJ,Marco2016},
\begin{equation}\label{eq:sigma}
\sigma = \frac{B_0^2/\mu_0}{n_0 m_e v_{\rm sh}^2},
\end{equation}
where $m_e$ is the electron mass, $n_0$ is the upstream density, and $v_{\rm sh}$ is the speed of the upstream plasma \emph{in the shock front frame}. The reduction of the density jump for a strong sonic shock $(P_{\rm upstream} = 0)$ is observed for $\sigma \gtrsim 1$ (see Figure 2 in \cite{Haggerty+2022}).

As it stands, the rationale is essentially macroscopic. We can therefore assume that it should remain valid for an electron-ion plasma, provided that $\sigma$ is rescaled as follows:
\begin{equation}\label{eq:sigmaei}
\sigma = \frac{B_0^2/\mu_0}{n_0 m_i v_{\rm sh}^2},
\end{equation}
where the only difference is that $m_i$, the ion mass, replaces $m_e$. We shall now verify this using PIC simulations.
\section{Particle-in-cell simulations}
\label{sec:setup}

\begin{figure*}[!ht!]
\includegraphics[width=18cm]{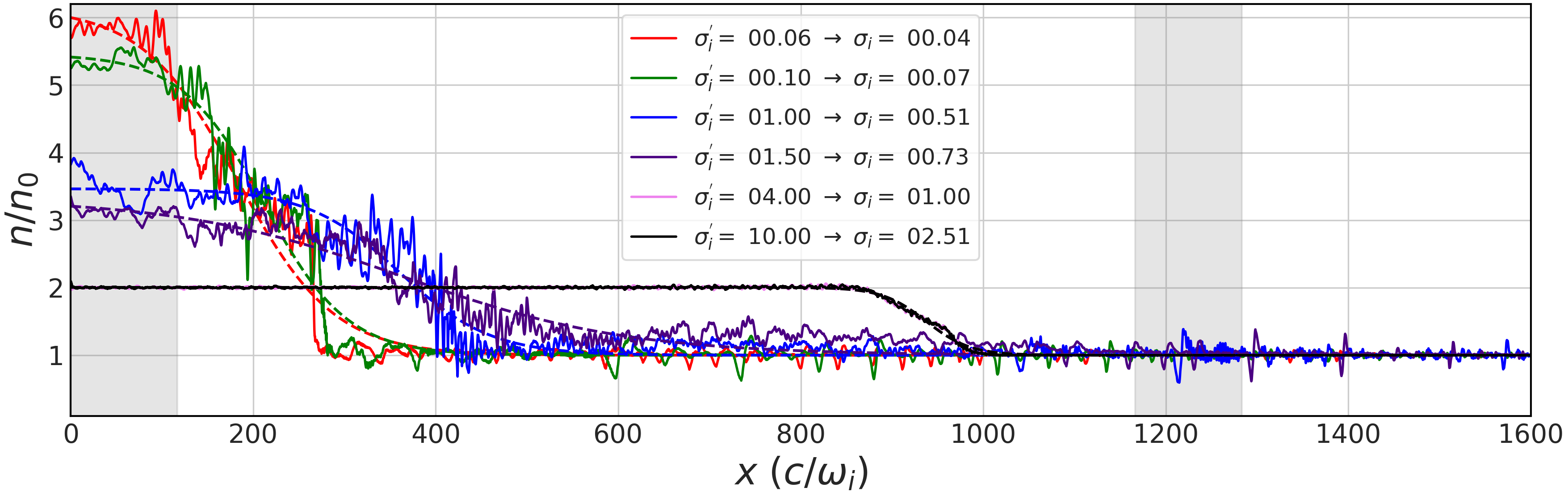}
\caption{%
\label{Fig:nden}%
The spatial distribution of the number density (electron+ions) in various simulations at $t \omega_i = 9400$. This shows that for values of $ \sigd > 4$ the compression ratio $R=n/n_0$ drops to $R \sim 2$ mean while it is $R>2$ for smaller values of $\sigd$.
Dashed lines for each simulation represent the function in Equation ~\eqref{Eq:ndenjump} using the best fit values of $R, \lambda_{\rm sh}$, and $x_{\rm sh}$ to number density profile.
When plotting the density profile in simulations (solid lines) we filter all fluctuations on scales smaller than $4c/\omega_i$ using a moving average filter. The gray-shaded regions, which extend over about $115\,c/\omega_i$, define the regions where we compute the numerical entropy density in simulations \add{(Figure \ref{Fig:R_S}, bottom panel)}.}
\end{figure*}

We conduct 1D3V (one spatial and three momentum dimensions) shock simulations using the Particle-in-Cell (PIC) code SHARP-1D3V \citep{sharp,sharp2}. The SHARP code demonstrates superior capability in avoiding numerical heating while maintaining exact momentum conservation. This feature is crucial for our work and is explicitly demonstrated in Section \ref{sec:results}. The code has been extensively utilized to study various phenomena, including beam-plasma instabilities \citep{resolution-paper,sim_inho_18,th_inho_20}, cosmic-ray-driven instabilities \citep{sharp2,Lemmerz+2023,Lemmerz2025}, and the formation of shocks in electron-ion plasmas \citep{Shalaby+2022ApJ,Shalaby2024ApJL}.

We employ a realistic ion-to-electron mass ratio, $m_r \equiv m_i/m_e = 1836$, where $m_i$ ($m_e$) is the ion (electron) rest mass. Additionally, we include a parallel large-scale magnetic field $B_0$, and all simulations are conducted in the downstream/contact discontinuity rest frame, following the setup used in \cite{Shalaby+2022ApJ}. While it is common practice to use a reduced ion-to-electron mass ratio in PIC simulations, \cite{Shalaby2024ApJL} has shown that a reduced mass ratio can significantly influence both thermal and non-thermal energy dissipation in parallel shocks across a range of Alfvénic Mach numbers. A reduced mass ratio can significantly alter the hierarchy of the instabilities involved in the shock formation \citep{BretPoP2010}. Therefore, we adopt the realistic mass ratio $m_r = 1836$ in this work.

For all simulations, the upstream average velocity for both electrons and ions is $v_u = -0.1c$, where $c$ is the speed of light. Consequently, the shock speed in the rest frame of the upstream plasma, $v_{\rm sh}$, depends on the shock compression ratio $R$ as follows:
\begin{equation}
v_{\rm sh} = v_u \left[ 1 + \frac{1}{R-1} \right] = \frac{R}{R-1} v_u.
\end{equation}

In all simulations, the upstream electron skin depth is resolved with 10 cells, and the number of particles is set such that the far-upstream region contains 200 particles per species (200 electrons and 200 ions) per computational cell. The time step is $0.045\,\omega_p^{-1}$, where $\omega_p = \sqrt{\omega_e^2 + \omega_i^2}$ is the total plasma frequency. The electron ($\omega_e$) and ion ($\omega_i$) plasma frequencies are given by:
$\omega_e = \sqrt{e^2 n_e/\epsilon_0 m_e} = \omega_p \sqrt{ m_r/(m_r + 1)}$, and
$\omega_i = \sqrt{ e^2 n_e/\epsilon_0 m_i } =  \omega_p/\sqrt{m_r + 1}  = \sqrt{m_r} \, \omega_e$, where $e$ is the elementary charge of the electron.

The magnetization parameter, representing the magnetic-to-kinetic energy ratio for ions in the shock, $\sigma_i$, is defined as:
\begin{equation}
\sigma_i \equiv \frac{B_0^2/\mu_0 }{n_0 m_i v_{\rm sh}^2} = \left( \frac{v_A^i}{v_{\rm sh}} \right)^2 = \frac{1}{\Ma^2},
    \end{equation}
where $n_0 = n_e = n_i$ is the average number density of electrons or ions in the far upstream region, $v_A^i$ is the ion Alfvén speed, $\mu_0 = 1/(\epsilon_0 c^2)$ is the permeability of free space, where $\epsilon_0$ is the vacuum permittivity, and $\Ma$ is the Alfvénic Mach number of the shock. This can also be expressed as:
\begin{equation}
\sigma_i = \frac{  B_0^2 /\mu_0}{n_0 m_i v_{u}^2} \left( \frac{R-1}{R} \right)^2 = \sigma_i^{'} \left( \frac{R-1}{R} \right)^2,
\label{Eq:SigR}
\end{equation}
where $\sigma_i^{'}$ is introduced for convenience.

In all simulations, we set a parallel (along $x$) constant magnetic field, which determines the value of $\sigma_i^{'}$. Since $v_u = -0.1c$ is fixed, $\sigma_i^{'}$ is varied by adjusting the value of  $B_0/\sqrt{n_0}$. We conduct six simulations with $\sigma_i^{'} = \{0.06, 0.1, 1, 1.5, 4, 10\}$. Note that the value of $R$ is not known a priori, and thus $\sigma_i$ is also not known in advance.

The initial conditions for particles are identical in all simulations. Electrons and ions are initially in thermal equilibrium with $k_B T_e = k_B T_i = 3.67 \times 10^{-5} m_e c^2$ with isotropic temperatures, i.e., $k_B T_{x} = k_B T_{y} = k_B T_{z} = 1.224 \times 10^{-5} m_e c^2$. This fixes the upstream sonic Mach number to
\add{$\Ms = v_{\rm sh}/\sqrt{2 \Gamma k_B T_i/m_i} = 387.3 \times R/(R-1)$, where $\Gamma = 5/3$ is the adiabatic index.}
Both electrons and ions in the far-upstream region drift toward the contact discontinuity at $x=0$ with speed $v_u = -0.1c$. We note that the temperatures  $T_i$ and $T_e$ match measured plasma values in the solar wind at $1$ au \citep[see, e.g.,][]{Maruca.etal:23}, whereas other plasma parameters, e.g., $n_0$ and $B_0$, or the shock speed exceed them by a few orders of magnitude, due to computational limitation. However, the $\Ma$ range covered here (and therefore the fast magnetosonic number \add{$M_f$}) matches the supercritical regime of interplanetary shocks at $1$ au \add{($M_f \gtrsim 1.5$ in low plasma beta regime for parallel shocks)}, making the simulations relevant to the interpretation of observations.

The far upstream plasma-$\beta$ for species $s$ (electrons or ions) is given by
\begin{eqnarray}
\beta_s
= 
\dfrac{ n_0 k_B T_s }{  B_0^2/2 \mu_0}
=
\dfrac{ n_0 m_i v_u^2 }{ B_0^2 /\mu_0 }
\dfrac{ 2  k_B T_s }{ m_i v_u^2}
=
\dfrac{ 1}{\sigd }
\dfrac{ 2  k_B T_s }{ m_i v_u^2}
\label{Eq:beta}
\end{eqnarray}
That is, the initial (upstream and isotropic) plasma $\beta$ for both electrons and ions in various simulations is given by  
\begin{eqnarray}
    \beta = 
\dfrac{    1.333 \times 10^{-6} }{\sigma_i^{'}}.
\end{eqnarray}

The simulation domain dynamically expands in the $x$-direction to prevent the escape of any accelerated electrons or ions.
At time $t \omega_i = 9400$, where we present results from various simulations in Section~\ref{sec:results}, the domain has expanded to $3000$--$4000\,c/\omega_i$ across different simulations.
The code also incorporates dynamical load balancing to ensure an even distribution of particles per core in both the upstream and downstream regions of the domain.

In the following sections, we only display the domain up to $1600\,c/\omega_i$  from the reflecting wall to focus on the shock transition regions.

\section{Simulation Results}
\label{sec:results}

In this section, we present the results of our simulations. We examine how the density jump, entropy jump, temperature jump, and particle acceleration (for both electrons and ions) vary as the system transitions to the strongly magnetized regime i.e., $\sigma_i \geq 1$.
\add{
While we present the final stages of our simulations in this section, the full dynamical evolution for all simulations can be found at \href{https://mohamadshalaby.github.io/Pshock_Sig01.html}{mohamadshalaby.github.io/Pshock\_Sig01.html}.
}

\subsection{Number density across shock transition}

The total \add{plasma} number density (for electrons and ions combined) profile  for various simulations at $t \omega_i = 9400$ is shown in Figure~\ref{Fig:nden}.
To characterize the density jump at various shocks, we fit the total number density across the shock with the following fit function (following \cite{Giacalone.Jokipii:09} and as done by~\citet{Haggerty+2022}) 
\begin{eqnarray}
\label{Eq:ndenjump}
\frac{n (x)}{n_0} 
= 
\frac{R+1}{2} 
+ 
\frac{R-1}{2} \tanh 
\left[ - \frac{x-x_{\rm sh}}{\lambda_{\rm sh}} \right]
\end{eqnarray}
Though nonlinear fitting, we find the best-fit values for shock compression $R$, shock half-width $\lambda_{\rm sh}$, and shock location, $x_{\rm sh}$.
The best fit values for $R$ is given in Table~\ref{tab:fitvalues} and the function using the fitted values for $R$, $\lambda_{\rm sh}$, and $x_{\rm sh}$ in various simulations is shown as dashed colored line in Figure~\ref{Fig:nden}.

\begin{table}[!ht!]
\centering
\footnotesize
\caption{\label{tab:fitvalues}%
Various simulations are initialized with different values for $\sigd$.  
We determine the compression ratio, $R$,  by fitting the number density (solid line in Figure \ref{Fig:nden}) to the  fitting function in Equation \eqref{Eq:ndenjump}. The resulting values of $\sigma_i$ and the \alf ic Mach number, $\Ma$, are derived using Equation \eqref{Eq:SigR} and the best fit value for $R$.  
The dependence of the compression ratio, $R$, on the inferred value of $\sigma_i$ is shown in the top panel of Figure~\ref{Fig:R_S}.%
}
\begin{tabularx}{\columnwidth}{Y|YYY}
\hline
$\sigd$   & $R$ & $\sig$  & $\Ma$  \\
\hline
\rule{0pt}{8pt}
$0.06$ & $6.15$  & $0.04$ & $4.88$ \\
$0.1 $ & $5.46$  & $0.07$ & $3.87$ \\
$1   $ & $3.46$  & $0.51$ & $1.41$ \\
$1.5 $ & $3.32$  & $0.73$ & $1.17$ \\
$4   $ & $2.0$   & $1.0$  & $1.0$  \\
$10  $ & $2.0$   & $2.51$ & $0.63$ \\
\hline
\end{tabularx}
\end{table}

The resulting compression ratios, $R$, for various simulations are also given in Table~\ref{tab:fitvalues}.
We then use Equation~\eqref{Eq:SigR}, to find the corresponding values for $\sigma_i$ and the ion {\alf}ic Mach number $\Ma$ and write down the returns in the last two rows of Table~\ref{tab:fitvalues}.
Moreover, the dependence of the compression ratio on the value of $\sigma_i$ is plotted in the top panel of Figure~\ref{Fig:R_S}.

This demonstrates that the compression ratio decreases to the expected value of $R \sim 2$ for strongly magnetized simulations ($\sigma_i > 1$).
At low magnetization, particularly for $\sigd < 1$, the downstream density exhibits large amplitude fluctuations (see the shaded region in Figure \ref{Fig:nden}), indicating that the downstream plasma has not reached a steady state at  $t \omega_i = 9400$. This results in best-fit compression ratios for $R$ significantly
larger than $4$.
At later times, when an approximate steady state is achieved, a higher compressibility ($R > 4$) is reached in  simulations due to the presence of accelerated particles \citep{Blondin.Ellison:01,Fraschetti.etal:10}, which is also seen in \add{other} kinetic scale simulations ~\citep{Shalaby+2022ApJ,Gupta2024}.

\subsection{Entropy across shock transition}

In this section, we study the change in entropy across \add{the transition region in various simulations}.
\add{Below, we explain how} we approximate the numerical values for the entropy. 

If we do not assume that the distribution in momenta is separable in each direction but rather depends only on the energy (similar to the case of an isotropic Maxwell-J\"uttner distribution in the rest frame), we can proceed as follows. If we divide the particle momenta, $u$, where $u= \gamma v$ is the spatial part of the 4-velocity, into momenta bins labeled with $k$ and of equal size $\Delta u$, we can write:

\begin{eqnarray}
&& f(u) \rightarrow  f_k = \frac{N_k}{4 \pi u_k^2 \Delta u},
\nonumber \\ 
&& \Rightarrow
\int d^3 u \, f(u) \rightarrow \sum_k 4 \pi u_k^2 \Delta u \, f_k = \sum_k N_k = N_p,
\end{eqnarray}
where $N_k$ is the number of particles with energies within the $k$th energy bin, $N_p$ is the total number of particles, \add{and all velocities are normalized by the speed of light, $c$.}  

Therefore, the normalized entropy density, $s$, is given by \citep{Bret+Ramesh2018,Bret2021}:
\begin{eqnarray}
&&
\frac{s}{k_B} =
\frac{- \int d^3 u \, f \ln f}{N_p },
\nonumber \\ 
&&
\Rightarrow 
\frac{s}{k_B}  \rightarrow 
 \sum_k 4 \pi u_k^2 \Delta u \, f_k \ln f_k
= 
\sum_k N_k  \ln 
\frac{N_k }{4 \pi u_k^2 \Delta u}.
\nonumber
\\
\label{Eq:NumS}
 \end{eqnarray}

To verify our algorithm, we compare it against the analytical entropy of an isotropic Maxwell-Jüttner distribution characterized by the normalized temperature, $T = k_B T_s / m_s c^2$. In this case, the distribution and the analytical entropy density are given by:
\begin{eqnarray}
f(\vec{u}) 
&=&
\frac{N_p}{4 \pi T K_2(1/T)} e^{-\gamma/T},
\label{Eq:MJdist}
\\
\frac{s}{k_B}
&=&
\ln \left[4 \pi T K_2\left( 1/T \right) \right]
+
\left[ 
 3 + \frac{K_1\left( 1/T\right)}{ T K_2\left( 1/T \right)}
\right]
-\ln N_p,
\qquad
\label{Eq:AnaS}
\end{eqnarray}
where $\gamma^2 = u_x^2 + u_y^2 + u_z^2 + 1$, and $K_1(x)$ and $K_2(x)$ are the modified Bessel functions of the first and second kind, respectively.  

To test our numerical approximation for computing the numerical entropy density, as described in Equation \eqref{Eq:NumS}, we initialize a sample of particle energies that follows the distribution in Equation \eqref{Eq:MJdist}. This is achieved using the algorithm described in detail by \cite{Zenitani2015}. We then compute the numerical entropy density using Equation \eqref{Eq:NumS} and compare it to the analytical expectation given by Equation \eqref{Eq:AnaS} for both relativistic ($T > 1$) and non-relativistic ($T \ll 1$) cases.  
The results show that our numerical algorithm is in excellent agreement with the analytical expectations when a sufficient number of particles and energy bins are used.

\begin{figure}
\includegraphics[width=8.2cm]{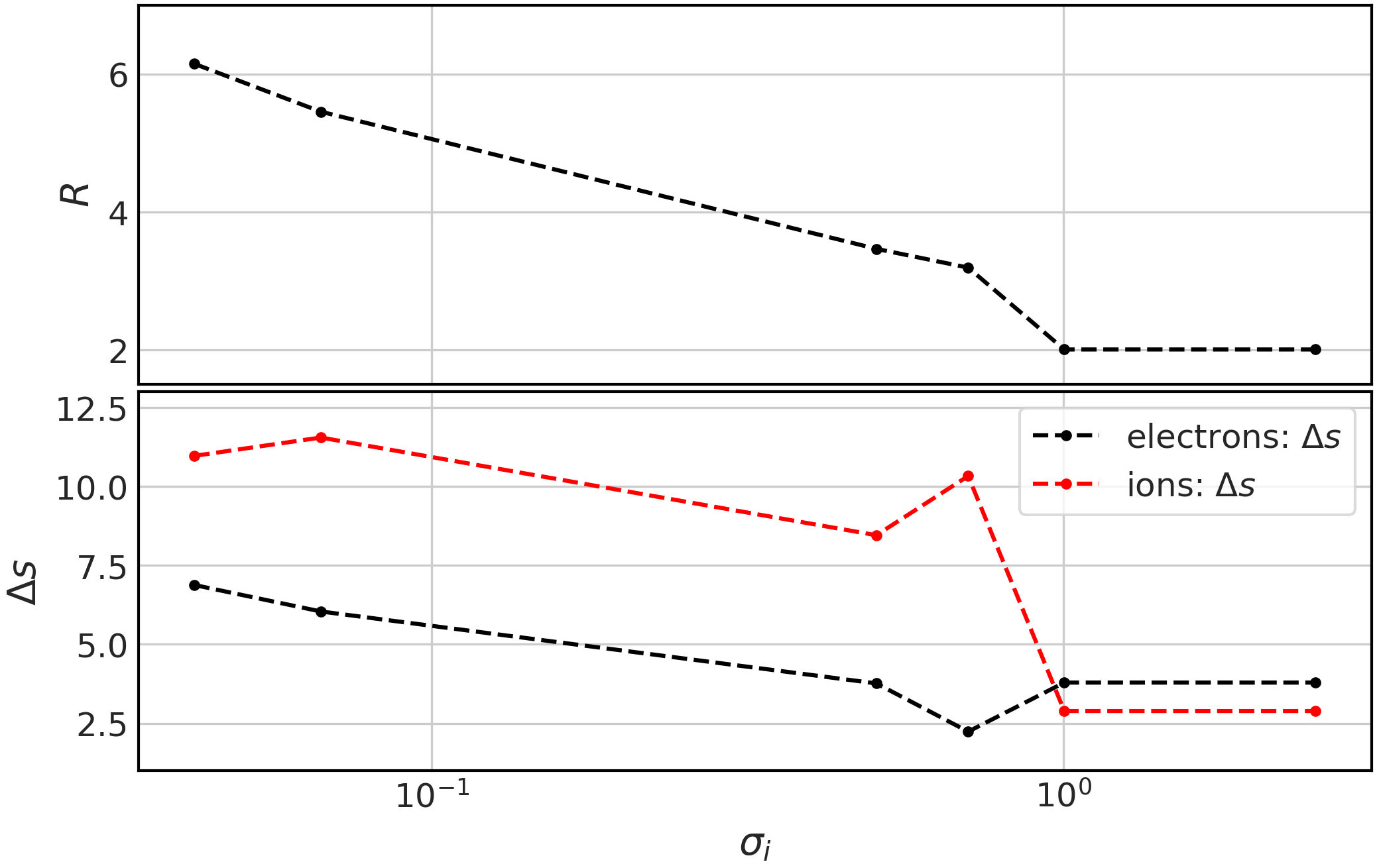}
\caption{%
\label{Fig:R_S}%
Top panel shows the dependence of the shock compression ration, $R$ on the value os $\sigma_i$.
Bottom panel shows the entropy jump between upstream and downstream regions highlighted with gray regions in Figure\ref{Fig:nden}, i.e., $\Delta s = s_{\rm downstream} - s_{\rm upstream}$.
In the bottom panel, dashed red (black) curves shows the entropy jump for ions (electrons) in various simulations.
This is shown at $t \omega_i = 9400$, the same time used in Figure \ref{Fig:nden}.
}
\end{figure}

Given the anisotropic nature of the plasmas in our simulations, we discretize the momentum space into a 2D grid with indices $k$ and $j$ representing the parallel and perpendicular directions, respectively. The grid consists of 1500 uniform bins in each direction, with equal spacing $\Delta u_{\parallel}$ and $\Delta u_{\perp}$. We then calculate the number of particles in each bin, $N_{k,j}$, and use this to approximate the numerical entropy as follows:
\begin{eqnarray}
\frac{s}{k_B}  \rightarrow 
\sum_{k,j} N_{k,j} \ln 
\frac{N_{k,j}}{2 \pi u^j_{\perp} \Delta u_{\parallel} \Delta u_{\perp}},
 \label{Eq:NumS2}
\end{eqnarray}
where, $u^j_{\perp}$ is the value of bin centers in the perpendicular direction. 
To compute the entropy jump across \add{the transition region in various simulations}, we calculate the entropy in the downstream and upstream regions, i.e., the gray-shaded regions in Figure~\ref{Fig:nden}, which extend over about $115\,c/\omega_i$ and use Equation~\eqref{Eq:NumS2} to compute the entropy densities numerically.
The entropy jump is then defined as $\Delta s = s_{\rm downstream} - s_{\rm upstream}$ for various simulations.  
The dependence of $\Delta s$ on the value of $\sigma_i$ is shown in the bottom panel of Figure~\ref{Fig:R_S} for both ions (red dashed) and electrons (black dashed) in various simulations.

For both electron and ion species, as shown in Figure~\ref{Fig:R_S}, an entropy jump is found in all simulations where the value of such jump is typically smaller in the strongly magnetized simulations with $\sigma_i \geq 1 $.
\add{
However, as shown in the next section, for $\sigd = 4$ and $10$, ions in the counter-propagating regions exhibit heating via electrostatic perturbations without achieving full thermalization. Consequently, the observed ion entropy jump in these cases stems solely from electrostatic heating rather than complete thermalization.
}

\subsection{Phase-space particle distribution across shocks}

\begin{figure*}[!ht!]
\includegraphics[width=18cm]{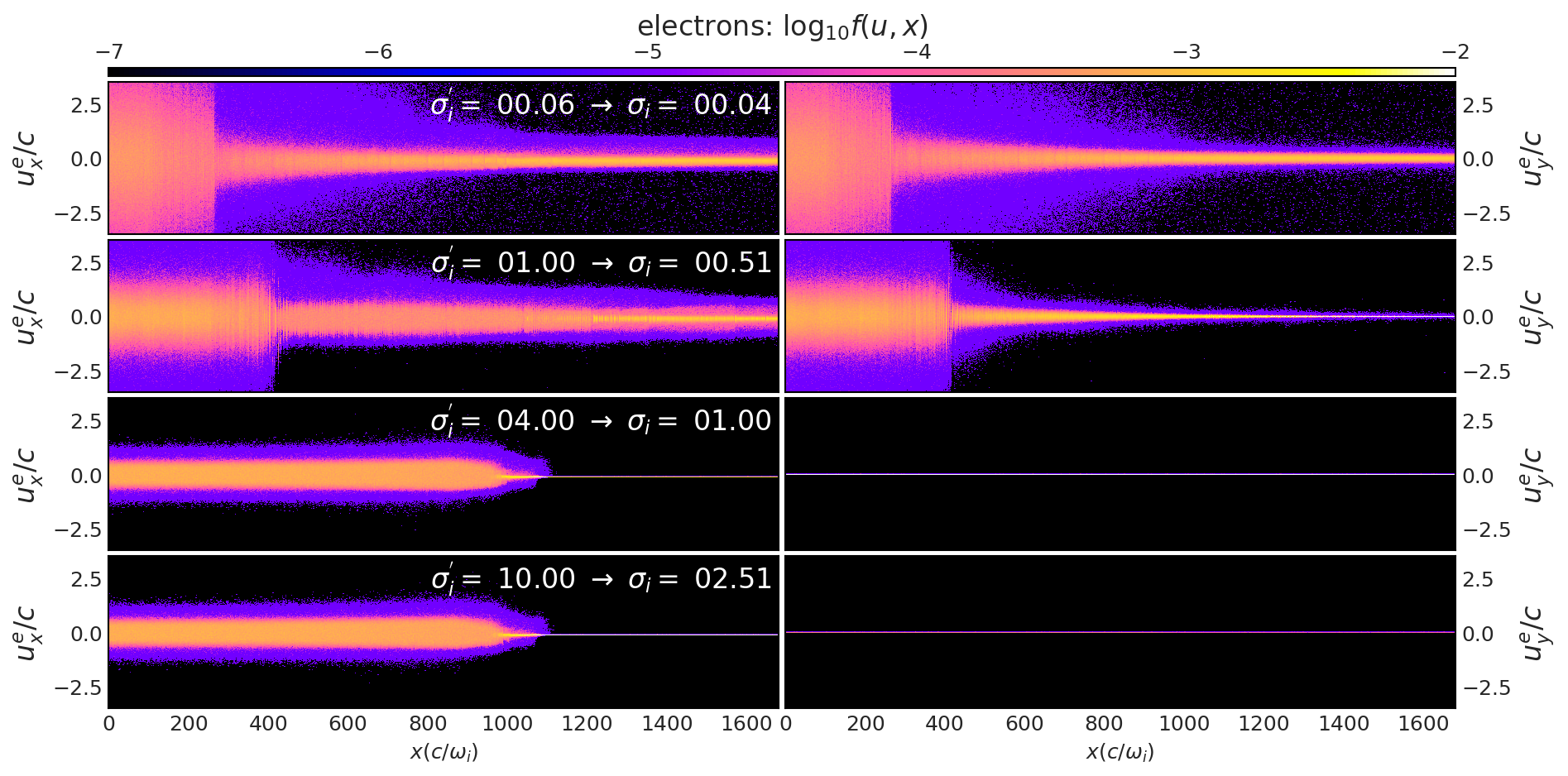}
\caption{%
\label{Fig:phase_lecs}%
Electrons phase-space distribution in various simulations at $t \omega_i = 9400$ (the same time used in Figure~\ref{Fig:nden}). The left panels show the spatial profile of parallel momenta, i.e., $f(u_x, x)$, while the right panels show that of perpendicular (y-component) momenta, i.e., $f(u_y, x)$.}
\end{figure*}

\begin{figure*}[!ht!]
\includegraphics[scale=0.57]{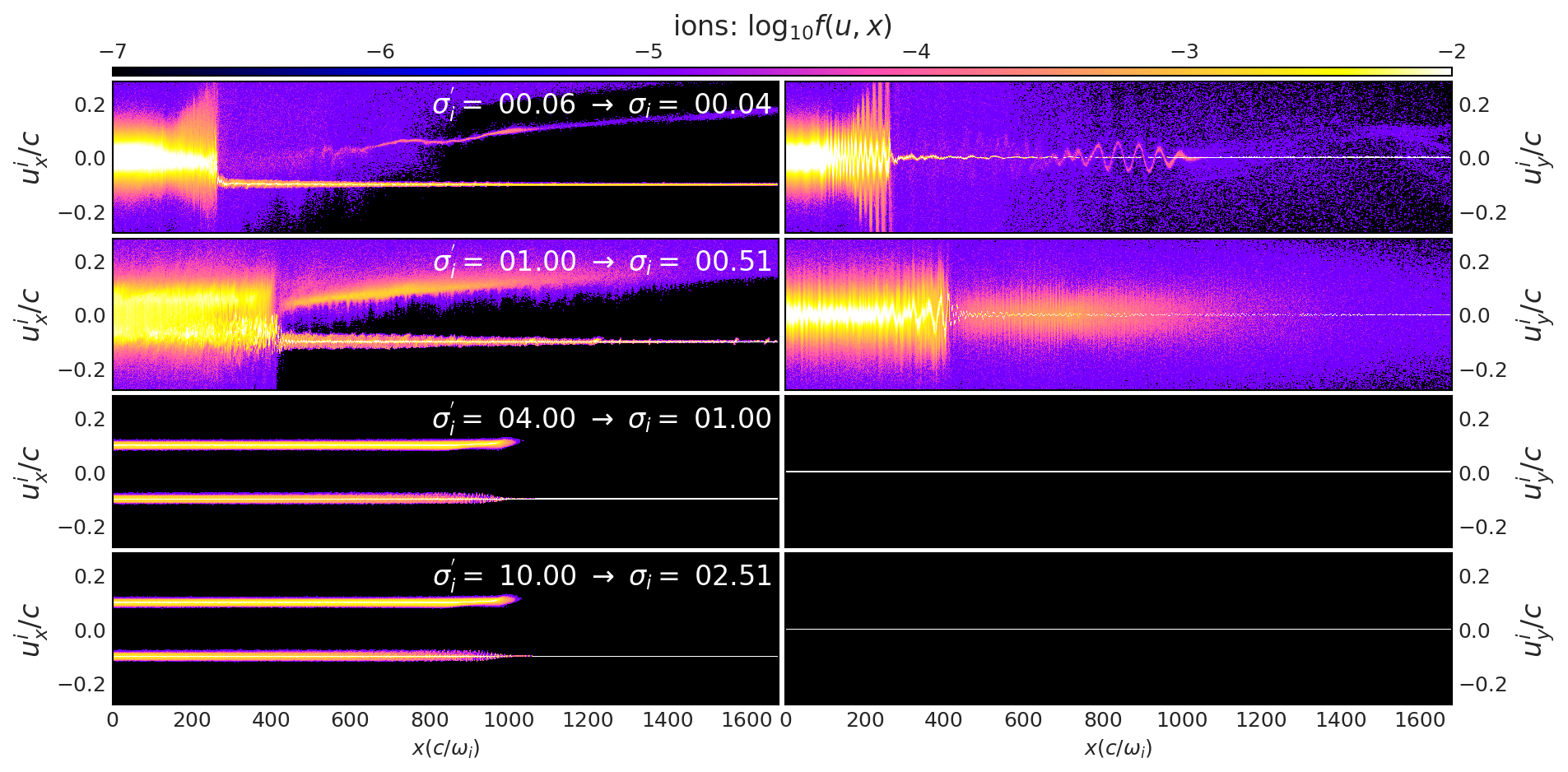}
\caption{%
\label{Fig:phase_ions}%
Ion phase-space distribution in various simulations at $t \omega_i = 9400$ (the same time used in Figure~\ref{Fig:nden}). The left panels show the spatial profile of parallel momenta, i.e., $f(u_x, x)$, while the right panels show that of perpendicular (y-component) momenta, i.e., $f(u_y, x)$.
}
\end{figure*}

In this section, we present the phase-space distribution of particles (electrons in Figure~\ref{Fig:phase_lecs} and ions in Figure~\ref{Fig:phase_ions}) in various simulations at $t \omega_i = 9400$.  
In both figures, we plot the 2D phase-space distribution for parallel momentum, i.e., $f(u_x, x)$, in the left column, where $u_j = v_j \gamma$ is the $j$ component of the particle 4-velocity, while the perpendicular momentum distribution, $f(u_y, x)$, is shown in the right column.  

In all simulations, the upstream kinetic energy of electrons is fully thermalized in the downstream region \add{(see Figure~\ref{Fig:phase_lecs})}. %
\add{This thermalization results from the well-known two-stream instabilities of the initially counter-streaming electron-ion beams \citep{Stix1992}. These instabilities destabilize the parallel (i.e., along $B_0$) electrostatic wave modes both at electron and ion skin-depth scales\footnote{\add{When focusing on parallel modes, the linear dispersion relation splits into two parts: one related to electrostatic instabilities (independent of the large-scale magnetic field $B_0$), and another containing only electromagnetic wave modes. See Appendix B of \citep{sharp2} for details}}.
In all simulations, these electric perturbations fully thermalize electrons in the beam interaction region on a very short time scale, $t\omega_i <10$.
However, this is insufficient to completely thermalize ions (regardless of the $\sigma$ value) as we show below.}

\add{Notably, the separation of the electron-to-proton scale by using herein the realistic $m_r$ allowed to temporally separate the downstream thermalization of the two species (seen also for protons at $\sigma<1$). In addition, since the upstream ram pressure is proton-dominated, in such an early stage the dissipation into heating is insufficient to form a stationary shock transition.
}

Parallel and quasi-parallel shocks that satisfy $1 < \Ma \lesssim \sqrt{m_r/R}/2$ can efficiently accelerate electrons~\citep{Shalaby+2022ApJ}. This is due to the destabilization of intermediate-scale unstable modes in these shocks~\citep{sharp2,Shalaby2023}.  
Indeed, for such shocks (top two rows in Figure~\ref{Fig:phase_lecs}), electrons are clearly accelerated and escape toward the upstream region of the shock.  
For the simulations with $\sigd = 4$ and $10$, electron acceleration is absent as expected.

A large velocity component along the shock surface, $u_y$, comparable to $u_x$, is observed in the two simulations with $\sigma_i < 1$ for both ions and electrons (top two rows of Figures \ref{Fig:phase_lecs} and \ref{Fig:phase_ions}). 
These two populations are likely to be efficiently energized into the supra-thermal range by shock drift acceleration via the motional electric field $|v_{\rm sh} \times B|$ (see section \ref{sec:ta})\add{, which is expected to accelerate particles efficiently at quasi-perpendicular shocks \citep{Jokipii:82,Decker:88,Krauss-Varban.Wu:89,Ball.Melrose:01,Fraschetti.Giacalone:15}}: a large upstream $B_\perp$ component becomes dominant over $B_{\parallel}$ in the downstream plasma. Since the oblique magnetic field persists past the shock ramp for hundreds of $c/\omega_i$, the \add{electron} population with large $u_y$ homogeneously fills the downstream region over a comparable length-scale.

An important consequence of this finding is that the \add{scattering of both ions and electrons into supra-thermal energies can be  
driven, or contributed, by magnetic fluctuations induced by temperature anisotropy at the shock \citep[as pointed out by][]{Quest.Shapiro:96}}, rather than by fluctuations pre-existing to the shock passage or by the streaming of resonant energized ions (which have not yet been generated). An in-depth analysis of this effect will be presented in a companion paper.
For shocks with $\sigma_i < 1$, while some ions and electrons become supra-thermal, most of the upstream kinetic energy is thermalized in the downstream region.


\add{%
For simulations with $\sigd = 4$ and $10$, at time $t \omega_i = 9400$, the driven electrostatic instabilities are insufficient to thermalize ions in the interaction region, as shown in the bottom two panels of Figure~\ref{Fig:phase_ions}. A similar configuration appears in the early stages of simulations with $\sigd < 4$, where ion thermalization occurs afterward concurrently with the destabilization of parallel electromagnetic wave modes, as shown in the top two panels of Figure~\ref{Fig:phase_ions}.
}

\add{%
Solving the electromagnetic dispersion relation for counter-streaming electron-ion beams yields a stability cutoff of $\sigd = 2$ for parallel electromagnetic wave modes \citep{Quest1988}. That is, parallel electromagnetic modes for counter-streaming ions become completely stable for $\sigd \geq 2$. This cutoff is independent of both the background (non-streaming) electron temperature and the counter-streaming ion plasma $\beta$ \footnote{\add{
To test this, we make use of the linear dispersion solver BO \citep{BO2_2021} which is publicly available at \url{https://github.com/hsxie/bo}.
}}.
For $\sigd < 2$ simulations, the unstable electromagnetic wave modes not only thermalize ions in the downstream region but also facilitate the transfer of part of the upstream ion kinetic energy to electrons, as shown in the top two panels of Figure~\ref{Fig:phase_lecs} and the top panel of Figure ~\ref{Fig:Temp}.
}

\add{%
It is important to note that oblique electromagnetic wave modes may become unstable at resonant wavelengths, potentially enabling complete ion thermalization for $\sigd > 2$. This possibility can be investigated through both linear analysis and ultimately 2D3V kinetic simulations; we defer this investigation to future work.
}

\add{For $\sigd >2$ cases,} we will see in the next section that the parallel ion temperature, in fact, increases by a factor of 100 (black dashed curve in the top panel of Figure~\ref{Fig:Temp}) in the downstream region compared to that of the far upstream due to these electrostatic unstable wavemodes.
In this work, we adopt the convention of referring to the denser part of the domain as the downstream region, regardless of whether it qualifies as a shock. This classification does not require the parallel ion momentum to be fully thermalized in the downstream region. Notably, in all simulations, the electrons are already fully thermalized in the downstream region.

\subsection{Temperature across shock transition}
\begin{figure*}
\includegraphics[width=18cm]{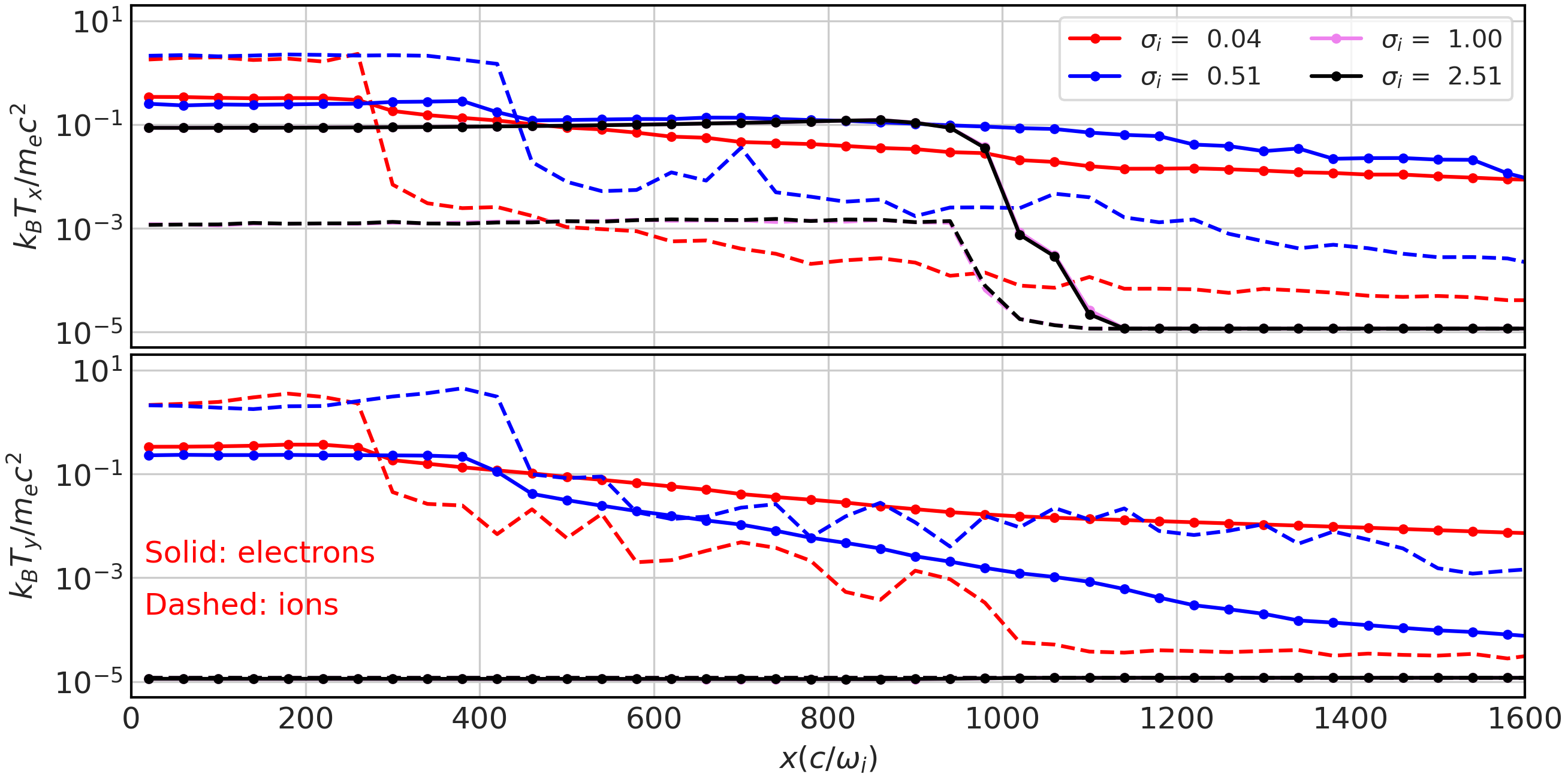}
\caption{%
\label{Fig:Temp}%
The spatial profiles for parallel (top) and perpendicular (y-component; bottom) temperatures in various simulations (color-coded) at $t \omega_i = 9400$ (the same time used in Figure~\ref{Fig:nden}).
In all cases, the ion temperature is represented by dashed curves, while the electron temperature is represented by solid curves with dots indicating the locations of bin centers where these temperatures are computed, and we use an equally spaced bins that extend over $40\,c/\omega_i$.%
}
\end{figure*}

In this section, we present the profiles for parallel and perpendicular temperatures (for electrons and ions separately) in various simulations. 
A robust method to compute the temperature of the thermal bulk of the distribution—particularly when the particle distribution includes non-thermal particles—is presented in the Appendix of \cite{Shalaby+2022ApJ}. This method relies on computing the particle energy spectra, $f(u)$, and inferring the temperature from the energy at which $u^4 f(u)$ is maximized for example. 

To avoid computing the full particle distribution at many location in the simulation when computed the spatial profile of the temperature, we compute the normalized  temperature, i.e., $k_B T/m_s c^2\rightarrow T$, using \citep{Wright+1975}

\begin{eqnarray}
T_j = \left\langle u_j v_j \right\rangle - \left\langle u_j \right\rangle \left\langle v_j \right\rangle,
\label{Eq:temp}
\end{eqnarray}

where $m_s$ is the species (ions or electrons) rest mass and the direction $j = x$ or $y$ or $z$.
The averaging in Equation \eqref{Eq:temp} is over all particles in some region of the simulation where the temperature is measured.

In the presence of non-thermal particles, this approach, however, could significantly overestimate the actual temperature of the thermal bulk of the distribution.
Therefore, we implement an iterative process where we determine $\left\langle u_j \right\rangle $ and $T_j$, and then include only particles where, $ u_j \in \left\langle u_j \right\rangle + 3 \times [\sqrt{T_j},-\sqrt{T_j}]$.
We then iteratively recompute $\left\langle u_j \right\rangle$ and $T_j$ from the filtered data. 
This procedure typically converges after 3-5 times iterations giving excellent agreement with the temperatures found from fitting the full particle distribution\footnote{We chose the interval  $3 \times [\sqrt{T_j}, -\sqrt{T_j}]$ so that, once non-thermal particles are excluded, more than 99 \% of the thermal particles are still incorporated. Increasing the pre-factor results in slower convergence.}.

From the geometry of the simulations, there is no preferred direction for the plasma streams in the perpendicular direction (see the right-side panels in Figure~\ref{Fig:phase_lecs} and Figure~\ref{Fig:phase_ions}). Therefore, the procedure outlined above can robustly compute the perpendicular temperatures for both ions and electrons.  
Moreover, as can be seen from the left-side panels of Figure~\ref{Fig:phase_lecs}, the parallel/kinetic energy is thermalized in the downstream region. Thus, this procedure also works well when computing the parallel electron temperature in various simulations.  

When computing the parallel temperature for ions, we observe from the left-side panels of Figure~\ref{Fig:phase_ions} that counter-streams exist throughout the downstream region in simulations with $\sigd = 4$ and $10$, with comparable densities. Thus, when computing the parallel ion temperature in the downstream regions, applying this procedure directly on all particles parallel momenta would yield a thermal spread of order $0.2c$, even though the thermal spread in individual streams is much smaller. Therefore, for these simulations ($\sigd = 4$ and $10$), we remove all particles with $u_x > 0$ before measuring the ion parallel temperatures, and thus we measure the ion temperature of the streams with $\left\langle u_x \right\rangle <0$ through out the simulation domain in such cases.

For $\sigd < 4$, parallel momentum is thermalized in the downstream region, and escaping/accelerated ions form a visible counter-stream, as seen in the top-left two panels of Figure~\ref{Fig:phase_ions}; our simulations show that the counter-stream speed settles to a uniform value at a later time ($t \omega_i \sim 12000$).
However, the density of the upstream escaping ions in these cases is much smaller compared to that of the incoming upstream ions (i.e., those with $u_x < 0$). Thus, when measuring the parallel ion temperature, our procedure, outlined above, always selects the incoming stream, and thus the temperatures shown in these cases correspond to that of the incoming stream.

It is important to point out that the temperature measured using Equation \eqref{Eq:temp} is the co-moving temperature, i.e., the temperature in the rest frame of the plasma \citep{Wright+1975}.
However, as can be seen from Figure~\ref{Fig:phase_lecs} and Figure~\ref{Fig:phase_ions}, the net drift of the thermal plasma bulk is non-relativistic. Thus, distinguishing between the temperatures in the plasma rest frame and those in the simulation frame is irrelevant.

The top and bottom panels in Figure~\ref{Fig:Temp} show the spatial profiles for parallel, $T_x (x)$, and perpendicular, $T_y (x)$, respectively, where various simulations are shown with different colors 
for both ions and electrons separately: ion temperatures are indicated by dashed lines, while electron temperatures are shown as solid lines. Dots highlight the locations of bin centers where these temperatures are computed. We use equally spaced bins when computing the temperatures that extend over $40\,c/\omega_i$ in all cases.

Numerical heating typically plagues PIC simulations, especially when simulating cold plasmas where the Debye length is not resolved \citep{sharp}.  
However, SHARP-1D3V is uniquely able to simulate such cases without resolving the cold-plasma Debye length and avoids such numerical heating by employing higher-order interpolation functions \citep{sharp,sharp2}.  
This unique ability of our PIC code is clearly demonstrated in the panels of Figure~\ref{Fig:Temp} after evolving the simulation for about $10^7$ time steps:  
The perpendicular temperatures are preserved at their initial values without any numerical heating in simulations with $\sigd = 4$ and $10$.
Even in other simulations where physical heating occurs, plasmas in the far upstream remain free of numerical heating in both parallel and perpendicular temperatures.

The top panel of Figure~\ref{Fig:Temp} reveals a jump in the parallel temperature across all simulations, whereas the bottom panel indicates that such a jump only occurs in simulations with $\sigma_i < 1$. In contrast, both electron and ion temperatures are well conserved for strongly magnetized shocks with $\sigma_i \geq 1$ \add{(the black and pink lines overlap)}.

Moreover, Figure~\ref{Fig:Temp} demonstrates that in weakly magnetized shocks with $\sigma_i < 1$, both ion and electron temperatures (parallel and perpendicular) are elevated in the shock precursor—the upstream region near the shock front—and gradually return to their initial values in the far upstream regions.

The bottom panel of Figure~\ref{Fig:Temp} demonstrates that the perpendicular temperature for both ions and electrons is conserved in strongly magnetized shocks ($\sigma_i \geq 1$). This conservation is a critical assumption often made in pair plasma calculations, e.g., \citep{Bret+Ramesh2018}.
Our simulations confirm that this assumption holds for both electron and ion species in the strongly magnetized electron-ion shocks as well.

\subsection{Temperature anisotropy}
\label{sec:ta}

\begin{figure*} 
\includegraphics[width=17cm]{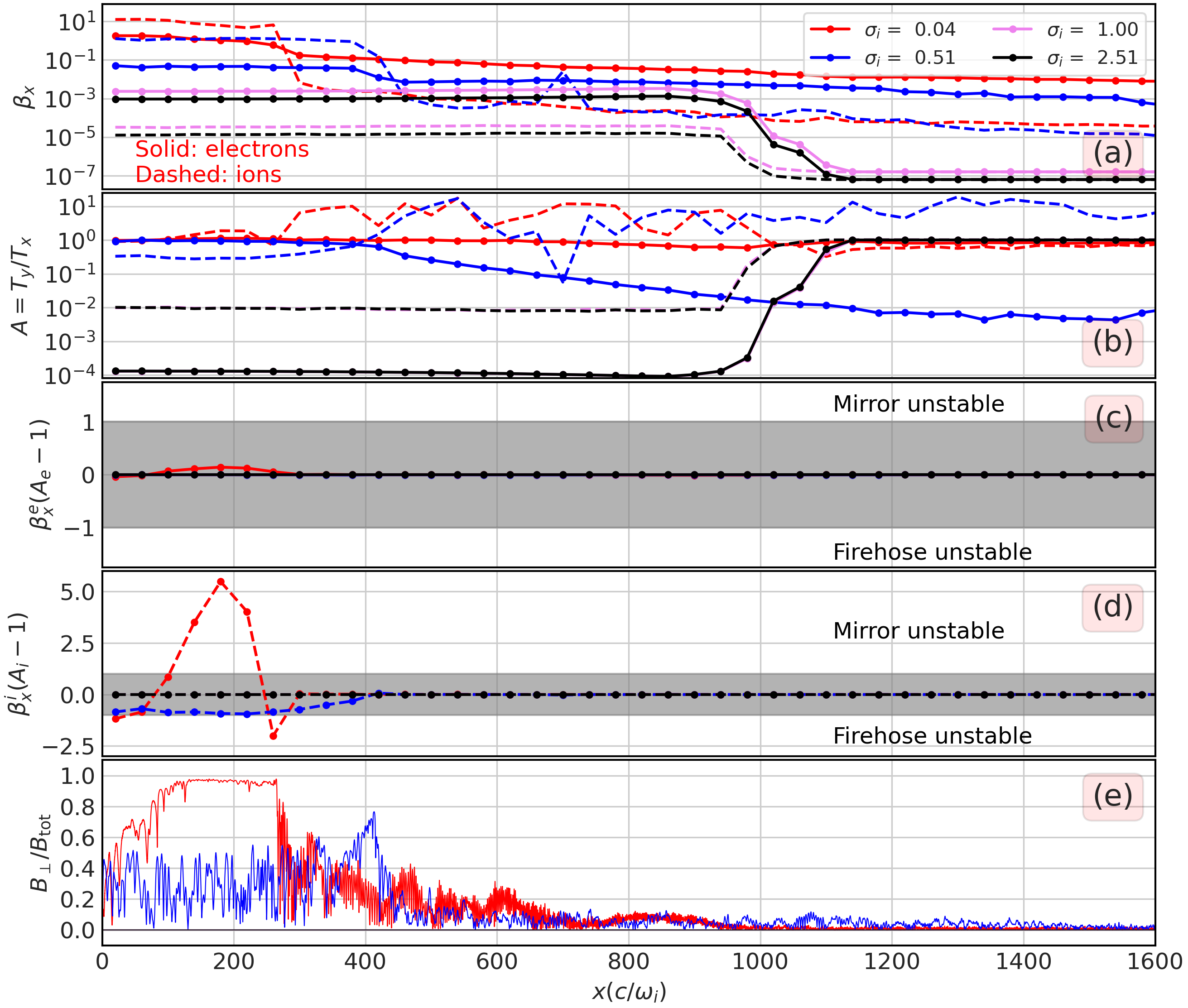}
\caption{%
\label{Fig:Stability}%
Spatial profiles of parallel plasma-$\beta$ (Panel a) and temperature anisotropy $A = T_y / T_x$ (Panel b) are shown for ions (dashed lines) and electrons (solid lines). Panel (c) demonstrates that electrons do not drive firehose or mirror instabilities, as $\beta_x (A_e - 1)$ lies within the gray stability region. 
Panel (d) reveals that for weakly magnetized shocks ($\sigma_i < 1$), ions drive mirror or firehose instabilities, while strongly magnetized cases ($\sigma_i \geq 1$) exhibit significant parallel ion heating without triggering any instabilities.
Panel (e) displays the spatial profile of the local magnetic field orientation, where $B_{\perp} = \sqrt{B_y^2 + B_z^2}$ and $B_{\rm tot} = \sqrt{B_x^2 + B_{\perp}^2}$. 
f
When plotting the perpendicular magnetic field profile in Panel (e), we filter out all fluctuations on scales smaller than $c/\omega_i$ using a moving average filter.
Dots on the solid lines indicate the locations of bin centers where the temperatures are computed, using equally spaced bins that extend over $40\,c/\omega_i$ in all cases.
All panels are shown at $t \omega_i = 9400$, the same time used in Figure \ref{Fig:nden}.
}
\end{figure*}

In this section, we analyze the level of temperature anisotropy in various simulations at time $t \omega_i = 9400$ and determine whether these anisotropy levels fall within the mirror or firehose instability regimes.

Panels (a) and (b) of Figure~\ref{Fig:Stability} display the spatial profiles of parallel plasma-$\beta$ and the temperature anisotropy parameter $A = T_y / T_x$, respectively. Ion profiles are represented by dashed lines, while electron profiles are shown with solid lines. Dots on the solid lines indicate the bin centers where the temperatures are computed. 

When computing the parallel plasma-$\beta$ profiles in panel (c), we use the definition in Equation~\eqref{Eq:beta} using local $T_x(x)$, and the number density calculated by computing the bin-averaged number density separately for ions and electrons. 
It is also important to note that we define the plasma-$\beta$ using the large-scale parallel magnetic field $B_0$, consistent with the definition in, e.g., \cite{Bret+Ramesh2018}, rather than the local (amplified in the downstream region) magnetic field.

The results reveal a prominent jump in both ion and electron (parallel) plasma-$\beta$ across all simulations, regardless of magnetization level. In the downstream region of strongly magnetized shocks ($\sigma_i \geq 1$), the level of anisotropy is significantly higher\add{, i.e., $A=T_{\perp}/T_{\parallel} \ll 1$,}
compared to simulations with lower magnetization. However, strongly magnetized \add{cases} exhibit a much smaller parallel plasma-$\beta$ for both electrons and ions.

\add{
When considering instabilities driven by ion and electron temperature anisotropies in our simulations,  we are restricted to instabilities in parallel electromagnetic wave modes only, as analyzed in, e.g., Sections 7.2 and 7.3 of \citep{Gary1993}.
We calculate spatial profiles of $\beta_x (A - 1)$ for both ions and electrons (Figure \ref{Fig:Stability}, panels c and d). 
Regions satisfying
$\beta_x (A - 1) > 1$ are mirror unstable\footnote{\add{
While temperature anisotropies with $A>1$ can excite oblique mirror modes (typically the fastest-growing electromagnetic instabilities are oblique), parallel modes remain present as subdominant features \citep{Gary1993}. 
In our simulations, we are necessarily restricted to these subdominant parallel mirror modes.}}, while that satisfy $\beta_x (A - 1) < -1$ are firehose unstable \citep{Gary1993,Bale.etal:09}. 
Regions where $-1 \leq \beta_x (A - 1) \leq 1$ are likely stable and are shown as the gray region in panels (c) and (d) of Figure \ref{Fig:Stability}.
}

As shown in Panel (c), electrons in all simulations remain within the stability regime, indicating that the anisotropy in electron temperatures does not drive mirror or firehose wave modes unstable.
The spatial profile of $\beta_x (A_e - 1)$ for electrons always lies within the stability region, denoted by the gray area in Panel (c).

For weakly magnetized shocks ($\sigma_i < 1$), ions are fully thermalized in the downstream region, and the resulting parallel and perpendicular temperature profiles indicate that the downstream plasma drives mirror or firehose ion-unstable wave modes. 
Notably, in the $\sigma_i = 0.51$ simulation, ions remain near the stability threshold for firehose instabilities throughout the downstream region.
As shown in Panel (e), firehose instabilities occur when the local magnetic field is quasi-parallel to the large-scale magnetic field ($B_0$) along $x$ \citep[see, e.g.,][]{Krauss-Varban.Omidi:91}, while the threshold for mirror-unstable wave modes is approached, for sufficiently large plasma-$\beta$ in the heated downstream plasma where the local magnetic field is quasi-perpendicular \citep[see, e.g.,][]{Violante.etal:95}. 
We also note an important difference in the upstream temperature anisotropy $A$: in the two simulations with $\sig < 1$, ions appear to fluctuate within mirror mode instability conditions ($A > 1$, see also panel (d)), while electrons fluctuate either within firehose conditions ($A < 1$ for $\sigma = 0.51$, see also panel (c)) or exhibit isotropy for $\sigma = 0.04$. We find that these fluctuations are not strong enough to drive upstream instabilities.

For strongly magnetized simulations ($\sigma_i \geq 1$), ions in the downstream region are not fully thermalized (see the bottom two panels of Figure~\ref{Fig:phase_ions}). However, they experience significant heating, with temperatures increasing by a factor of $100$ compared to the incoming ions. This jump in the parallel ion temperature—while the perpendicular temperatures remain conserved at their initial values—creates a large temperature anisotropy. Nevertheless, this anisotropy in ions is insufficient to drive firehose or mirror modes unstable.

\section{Conclusions}
\label{sec:conclusions}

In this paper, we investigate the impact of strong magnetic fields on collisionless electron-ion shocks using Particle-in-Cell (PIC) simulations. We explore how varying the ion magnetization parameter $\sigma_i$ affects shock jump conditions, \add{for} density, entropy, and temperatures \add{separately for both species}.

We summarize our key findings below:
\begin{itemize}
    \item Density Compression: Strongly magnetized \add{cases} ($\sigma_i > 1$) exhibit a lower compression ratio ($R \sim 2$), while weakly magnetized shocks show $R > 2$. At low magnetization, downstream density variations \add{(large-amplitude fluctuations)} indicate simulations have not reached steady state.
    
    \item Entropy increase: An entropy jump occurs in all \add{cases}, with smaller magnitudes in strongly magnetized simulations ($\sigma_i \geq 1$) for both electrons and ions.
    
    \item Particle Acceleration: In weakly magnetized shocks ($\sigma_i < 1$), ions and electrons are heated to a supra-thermal population, with most upstream kinetic energy thermalized downstream. In strongly magnetized \add{cases} ($\sigma_i = 4$ and $10$), electron acceleration is absent, and full electron thermalization occurs downstream. Ion thermalization remains incomplete at $t\omega_i = 9400$.
    
    \item Temperature Anisotropy: Strongly magnetized cases maintain pressure anisotropy stability downstream, leading to low compression ratios and preventing particle acceleration. Anisotropy is significantly higher downstream for $\sigma_i \geq 1$ due to conserved perpendicular temperature and increased parallel temperature.
    
    \item Plasma-$\beta$ and Instabilities: A jump in ion and electron (parallel) plasma-$\beta$ is observed in all simulations. Weakly magnetized shocks ($\sigma_i < 1$) drive mirror and/or firehose ion-unstable wave modes both upstream and downstream, whereas
    strongly magnetized cases ($\sigma_i \geq 1$) suppress these instabilities in both species.
\end{itemize}
The simulations confirm that the assumption of conserved perpendicular temperature holds for both electrons and ions in strongly magnetized simulations. The SHARP-1D3V code effectively avoids numerical heating, even for cold plasmas, which is essential in demonstrating these results.

\section*{Acknowledgments}
This work was supported in part by Perimeter Institute for Theoretical Physics.  Research at Perimeter Institute is supported by the Government of Canada through the Department of Innovation, Science and Economic Development Canada and by the Province of Ontario through the Ministry of Economic Development, Job Creation and Trade.
M.S. receives additional support through the Horizon AstroPhysics Initiative (HAPI), a joint venture of the University of Waterloo and Perimeter Institute for Theoretical Physics.
A.B. acknowledges support by the Ministerio de Economía y Competitividad of Spain (Grant No. PID2021-125550OB-I00).
F. F. was partially supported by NASA under grants 80NSSC21K0119, 80NSSC24K1240. Partial funding for this project was supplied by the PSP project through the SAO/SWEAP subcontract 975569.
This work was supported by the North-German Supercomputing Alliance (HLRN) under projects bbp00046 and bbp00072. 

\bibliography{refs}
\bibliographystyle{aasjournal}

\end{document}